# Structural Textile Pattern Recognition and Processing Based on Hypergraphs

**Vuong M. Ngo · Sven Helmer · Nhien-An Le-Khac · M-Tahar Kechadi**

**Abstract** The humanities, like many other areas of society, are currently undergoing major changes in the wake of digital transformation. However, in order to make collection of digitised material in this area easily accessible, we often still lack adequate search functionality. For instance, digital archives for textiles offer keyword search, which is fairly well understood, and arrange their content following a certain taxonomy, but search functionality at the level of thread structure is still missing. To facilitate the clustering and search, we introduce an approach for recognising similar weaving patterns based on their structures for textile archives. We first represent textile structures using hypergraphs and extract multisets of $k$-neighbourhoods describing weaving patterns from these graphs. Then, the resulting multisets are clustered using various distance measures and various clustering algorithms (K-Means for simplicity and hierarchical agglomerative algorithms for precision). We evaluate the different variants of our approach experimentally, showing that this can be implemented efficiently (meaning it has linear complexity), and demonstrate its quality to query and cluster datasets containing large textile samples. As, to the best of our knowledge, this is the first practical approach for explicitly modelling complex and irregular weaving patterns usable for retrieval, we aim at establishing a solid baseline.

Vuong M. Ngo
Ho Chi Minh City Open University, 35-37 Ho Hao Hon, Dist. 1, HCMC,
Vietnam E-mail: vuong.cs@gmail.com or vuong.nm@ou.edu.vn

Sven Helmer
Department of Informatics, University of Zurich, Binzmühlestrasse 14, Zurich,
Switzerland E-mail: helmer@ifi.uzh.ch

Nhien-An Le-Khac, M-Tahar Kechadi
School of Computer Science, University College Dublin, Belfield, Dublin 4,
Ireland E-mail: an.lekhac@ucd.ie, tahar.kechadi@ucd.ie

# 1 Introduction

In the humanities, the digitisation of cultural heritage and cultural practices plays an ever more important role (Stone 2012; Rosner et al. 2014). However, digitising cultural heritage is far from trivial. In many cultures textiles play a prominent role (Schoeser 2012), for instance in African (Clarke et al. 2015), Andean (pre-Columbian) (Bjerregaard and Huss 2017), British (Gale et al. 2012), Chinese (Kuhn 2012), Greek (Spantidaki 2016), and Indian (Fotheringham 2019) civilisations. They were used to communicate information, such as social standing, and are therefore important for researchers studying the history and prehistory of these regions. The traditions around weaving and the creation of textiles are kept alive by communities throughout the whole world, located in countries such as the United Arab Emirates, China, Vietnam, and Peru. However, this is becoming more and more difficult, as there is a waning interest in these traditions among younger generations, which makes the preservation of this cultural heritage a timely and pressing issues.

There are a number of digital archives for textiles that are publicly accessible, but these archives offer limited functionality when it comes to searching the collections. For example, the TEXMEDIN digital library (http:// www.texmedindigitalibrary.eu/) provides keyword-based search facilities. The Textile Museum of Canada (http://www.textilemuseum.ca/) goes fur- ther by allowing users to browse the collections according to different cate- gories, such as textile type, region, materials, techniques, and period and the University of Leeds International Textile Archive (ULITA) (http://ulita. leeds.ac.uk/) organises their collection by region. In the context of an ear- lier project (Brownlow et al. 2015; Martins et al. 2013), called "Weaving Com- munities of Practice", we went even further by constructing an ontology for Andean textiles and utilising this ontology for building a knowledge base, of- fering additional querying functionality. All of the platforms above require the use of certain terminology or keywords to make them work, though. During several visits to South-American museums taking place in our earlier project, the domain experts encountered new textile patterns previously unknown to them, which means that in some cases the exact terminology to describe these textiles is still missing. This motivated us to come with an approach to let the textiles speak for themselves, i.e., developing a method to compare textile patterns directly without first creating a (natural) language description.

While there are formal mathematical models for representing very regularly shaped grid-like textile patterns produced by machines, it is much harder to model manually created textiles, which exhibit a much more complex and irregular internal structure. We were not able to find an efficient technique powerful enough to represent the patterns we were confronted with in Andean and Vietnamese textiles. We opted for a hypergraph-based model: (graphs and) hypergraphs have been used widely to represent human-made objects, in fields as diverse as knowledge bases, natural language and image representation, and medicine. What made hypergraphs particularly interesting for us is the fact that they have been successfully applied to represent and model objects, their contexts, and the spatial relationships of subcomponents (Wong et al. 1989). After an ineffective attempt to model complex and irregular textile patterns with labelled regular graphs, we found hypergraphs to offer the functionality and expressibility we were looking for.

The purpose of this paper is to propose a hypergraph-based model for textile representation and develop suitable methods on the hypergraphs for textile pattern retrieval and clustering. This would not only help in the search and recognition process, but would also allow domain experts to gain deeper insights by being able to quantify differences and variations in patterns that evolved over time and in different regions. In summary, we make the following contributions here:

- We develop a novel approach based on hypergraphs for representing textiles. Our approach can handle many different structures, like woven, knitted, or braided textiles and it is invariant to orientation. To the best of our knowledge, this is the first practical approach that can handle very complex textile patterns.
- We propose a 2-step approach to measure the structural similarity of textiles. First, multisets of $k$-neighbourhoods, which describe the weaving structures from the hypergraph representation, are extracted. Essentially, these neighbourhoods are star-shaped subgraphs of hypergraphs. In a second step, the multisets are compared through various distance measures.
- We show the efficiency and effectiveness of our technique in querying and clustering a data set of $1,600$ textile samples, measuring the performance of our similarity measure. We validate the results obtained in our earlier work (Helmer and Ngo 2015) by evaluating our approach under multiple different scenarios, utilising a larger and more diverse data set and new distance and quality measures. The results we get back up our earlier results and demonstrate the robustness and generality of our approach. Our aim is to establish a baseline for modelling textile structure usable for identification and retrieval of weaving patterns.

The remainder of the paper is organised as follows. In the next section we review the related work. Section 3 covers the state-of-the-art for modelling textile structures and discusses their advantages and disadvantages. The new approach is introduced and detailed in Section 4. The new similarity measures were applied to well-known and popular unsupervised learning (clustering) algorithms in Section 5. We evaluate our approach experimentally: we present the methodology in Section 6 and the results in Section 7. We conclude and give some future directions in Section 8.

## 2 Related Work

2.1 Terminology-based Approach

A widely-used terminology for textiles and the basic patterns they are made of was compiled by Emery (2009), who, at the time of writing, was a curator at the Textile Museum in Washington D.C. We provide more details on Emery's classification and particular issues in Section 3. Although the terminology is not always completely consistent (Brezine 2009), it is a comprehensive work that systematically classifies textiles according to their internal structure. Nevertheless, it has some gaps when it comes to textiles created in various cultural contexts, including the South American Andes (Arnold and Dransart 2014; D'Harcourt 2002). It is also very challenging to try to find a natural language description for every possible textile structure, since there is a large diversity of textile patterns, especially when looking at manually created fabrics. Thus, it comes as no surprise that researchers have tried to develop formal and mathematical models to describe textile structures (Grishanov et al. 2009a,b).

2.2 Topology-based Approach

When reviewing formal models, we have to distinguish between two different types: those for regular grid-like structures and those for more irregularly shaped patterns. Mechanical looms create very regular patterns, which can be represented with the help of matrices (Milasius and Reklaitis 1988). As these methods are not adequate for describing complex patterns in handcrafted textiles, we focus on the second type of models. Topology-based approaches used elements from knot theory to describe textile patterns. Grishanov et al. (2009a,b) went further by developing a method using tangles, i.e., knot fragments embedding arcs into a sphere.[1] Although this is a more generally applicable approach, it still has some drawbacks. It can only be applied to structures that show periodicity (in two perpendicular directions) and it does not consider multi-layered disjoint textiles. Additionally, the topology-based models focus on the classification and enumeration of textile patterns, while we are interested in their fast retrieval. However, checking the equivalence of two structures made up of knots, links, or tangles is intractable in the general case (Cromwell 2004).

2.3 Textile Image-based Approach

We now turn to a completely different approach: describing textile patterns not with the help of abstract models, but with images taken with cameras. Many papers exploiting supervised learning techniques have been applied to defect detection (Yapi et al. 2015; Li et al. 2019a), fabric classification (Jing et al. 2019; Arora et al. 2019), and textile retrieval (Deng et al. 2018; Xiang et al. 2019). However, these approaches rely on textile colour rather than structure and share some general drawbacks of supervised learning techniques, such as being heavily dependent on the training data sets. Some researchers exploit the regularity of textures for material inspection (Ngan and Pang 2009), density detection (Zheng and Wang 2017), and textile recognition (Chan et al. 2017), while others use Fourier transform for the retrieval of weavings (Zhang et al. 2019) or clustering fabrics (Zhang et al. 2017). There is also work on utilising an entropy-based method to calculate the distribution of weave points (Zheng and et al. 2009). Applying image processing to the analysis of textile structures has the advantage that the process can often be automated. Most of the approaches described above are not generally applicable, though, i.e., they can only be used for specific patterns, such as knitting or regular grid-like structures. Additionally, since textiles are three-dimensional objects, some features may be hidden and many image-processing approaches have problems with this. While Ma et al. were able to extract some of the hidden information (Ma et al. 2011), they can only do so for regular grid-like structures.

---

[1] We provide more background on knots, links, and tangles in Section 3.

Varma and Zisserman (2009) propose texton-based representations, obtained from single images under unknown viewpoints and illumination, for fabric classification. These representations are suitable for modelling compact neighbourhood distributions with Markov random fields. Xie et al. (2015) use a two-step texton-encoding algorithm to classify the whole image, using a learned dictionary and the corresponding sparse coefficients over the features extracted from the image. Li et al. (2019b) propose a low-rank representation that divides an image into blocked matrices for dimensionality reduction with the goal of detecting various fabric defects. They also apply Eigen-value decomposition on blocked matrices to implement a version without a training phase. Kang and Zhang (2019) devise an Elo-rating algorithm of integral images to improve fabric products and to speed up the detection of defects. It needs to be trained on defect-free samples of a particular textile pattern to be effective. This means, that it cannot be used for categories that do not have an unambiguous representative. The categories of complex textile patterns we look at do not necessarily have unique representatives. Additionally, all the methods we just discussed work on images of textiles, not directly on their structure.

## 2.4 Application of Graph and Hypergraph Representation

Graphs and hypergraphs have many applications in various fields. For instance, they are used to represent structure of both natural and human-made objects. The NLP researchers apply graph for constructing a knowledge base to retrieval complex text answer (MacAvaney et al. 2019; Sawant et al. 2019) or integrate entity and concept (Shalaby et al. 2019). Qiao et al. (2020) model location-based social networks by using heterogeneous graphs which describe representations of users, points of interests, and temporal information. Gbadouissa et al. (2020) propose a heuristic clustering based on hypergraph theory, which optimises and effectively manages the used energy of sensor nodes in wireless sensor networks. Lierde and Chow (2019) use hypergraph transversals for text summarisation. In that context, nodes are the sentences of the corpus and hyperedges are themes which group sentences belonging to the same topics. Liang et al. (2019) propose hypergraphs to model electroencephalography (EEG) signals for emotion recognition. Here, a vertex is one trial of EEG signals, and the relationships among the vertices are the extracted EEG characteristics in the three domains of frequency, time, and wavelets. However, so far there is no work using graphs or hypergraphs for modelling textile structures.

## 3 Textile Modelling in the State-of-the-Art

Before presenting our approach, we define some basic concepts currently used for modelling textiles and also discuss their merits. One of the most important is the notion of *textile structure*, which describes the spatial relationships between segments or pieces of fibre, e.g. yarns, threads, or strands. Emery distinguishes three different general structures found in textiles (Emery 2009): felted fibres and interworked and interlaced elements. The first one, felted fibre, is of no particular interest to us, as it describes textiles in which fibres are compressed, matted, and condensed, resulting in dense and very irregular entanglements. This makes it very hard or almost impossible to identify the relationships between individual fibres. However, this is not an issue, since felted fibres are not compared to each other on this level of detail, as the individual entanglements will look different even for the same type of textile. The remaining two structures, interworked and interlaced elements, are much more important to us. In the former, techniques such as knotting, linking, stitching, looping, or twining are used to connect threads; in the latter, threads pass over and under each other (and this is the only way they connect). Figure 1 shows typical examples for interworked ((Fig. 1(a) and (b)) and interlaced elements (Fig. 1(c) and (d)). Here we offer only a short glimpse into Emery's classification scheme, giving a comprehensive overview would be beyond the scope of our paper. Especially, because we are not interested in providing natural language description, but we want to develop a formal model. We now turn to the topic most important for us: given (a fragment of) a textile pattern, find other textiles the the same or a similar arrangement of the constituent elements.

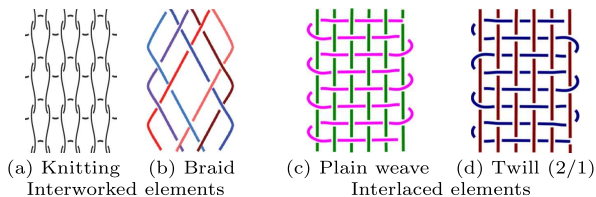

(a) Knitting    (b) Braid    (c) Plain weave    (d) Twill (2/1)
Interworked elements              Interlaced elements

Fig. 1: Examples of textile structures

### 3.1 Knots

In the following we give a brief summary of topological concepts taken from (Cromwell 2004; Grishanov et al. 2009a) that have been used to model textile structures, starting with the concept of *knots*, which are one-dimensional subsets of points $K \subset \mathbb{R}^3$ homeomorphic to a circle. A trivial knot, a *circle*, is depicted in Figure 2(a), while a more complicated structure, a so-called *trefoil* is shown in Figure 2(b). One important way to compare knots is to check whether they are equivalent or not. An intuitive notion of equivalence asks if we can transform one knot into another one by deforming it without breaking or cutting it. Two knots that can be transformed into one another are depicted in Figures 2(c) and (d).

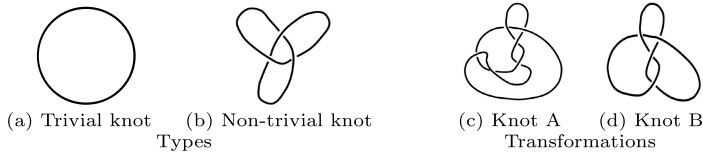

(a) Trivial knot    (b) Non-trivial knot         (c) Knot A     (d) Knot B
            Types                                  Transformations

Fig. 2: Examples of knot types and transformations

A well-known topological deformation is a *homotopy*, which is a continuous mapping of a space $X \subset \mathbb{R}^3$ over time. However, a homotopy is not sufficient to accomplish the task at hand, i.e., checking whether a continuous deformation is possible. In order to distinguish knots we need the concept of *ambient isotopy*. Rather than deforming the subspace $X \subset \mathbb{R}^3$, we distort the whole space around $X$, carrying it along.

### 3.2 Links

A generalisation of a knot is called a *link*, which is a set of entangled knots; Figure 3(a) illustrates a *trivial link* that we get by untangling or unlinking the knots to obtain a simple reference link. A more complicate structure, called *Borromean rings*, is shown in Figure 3(b). Similar to knots, the equivalence of links can be defined using ambient isotopy, i.e., we can check whether a link can continuously be deformed into another link without breaking or cutting any knots.

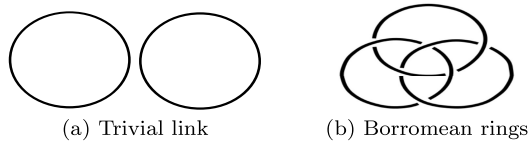

(a) Trivial link         (b) Borromean rings

Fig. 3: Examples of links

### 3.3 Tangles

Knot theory is a well-established mathematical field, but we found that we could not apply it directly to our use case, since textiles are not created by combining and entwining circles. Nevertheless, the notion of a *tangle*, introduced by Conway (1970), is much closer to what we need. Tangles are fragments of knots and Conway tried to simplify the enumeration and classification of knots and links with the help of tangles. It turns out that tangles are also useful when it comes to describing textiles: Grishanov et al. first applied them to describing textile structures (Grishanov et al. 2007).

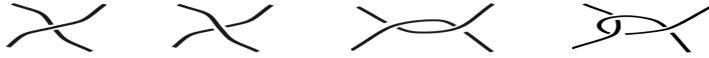

Fig. 4: Examples of tangles

Originally, Conway defined a tangle as a fragment of a knot with arcs ending in the four corners, which are labelled NW, NE, SW, and SE (like a compass rose). For examples of tangles, see Figure 4. The original definition of 2-tangles, containing two disjoint arcs and a collection of loops, can easily be extended to n-tangles, containing $n$ disjoint arcs. In our approach, we basically use simple 2-tangles containing exactly one crossing. We break down more complex structures into simple 2-tangles. We would represent the first two structures in Figure 4 directly, while breaking down the third structure into two tangles and the fourth one into three tangles.

3.4 Two-dimensional Projections

Although strictly speaking textiles are three-dimensional objects, for an abstract description and classification a two-dimensional representation is sufficient in most cases. We do not know the exact height of a point in a two-dimensional representation, but we can still clearly distinguish different types of textile patterns, as textiles do not protrude far into the third dimension. Mathematically, we are projecting knots, links, and tangles from $\mathbb{R}^3$ to $\mathbb{R}^2$. We have to be careful when projecting these elements, though. First, an edge is not allowed to be parallel to the projection direction. Second, the projection needs to be regular, i.e., it is injective except for a finite number of points. These exceptions are crossings in a link and we may only have at most two points in a link projected onto a crossing. Third, we have to be able to distinguish the arc on top from the one below. Usually, the lower arc is represented by a break in the line, a convention which we have used implicitly so far (and will keep using).

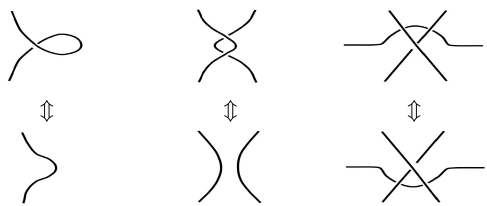

Fig. 5: Reidemeister moves

A two-dimensional projection creates some problems when transforming links by gradually deforming them, though. A transformation that is perfectly fine in $\mathbb{R}^3$ can lead to a situation in which the projection to $\mathbb{R}^2$ is not regular anymore. Reidemeister identified the problematic cases and defined his *Reidemeister moves*, shown in Figure 5, that allow a transformation to avoid the critical transformation steps by skipping over them. When checking the equivalence of two links, we need to find a sequence of gradual deformations and Reidemeister moves that transform one link into the other.

3.5 Discussion

Researchers such as Grishanov were interested in determining the equivalence of textile structures with the help of topological methods. Mapping this problem to knot theory allows the application of methods developed for determining the equivalence of knots. While this works on a theoretical level, in practice the situation is much more complicated. One of the first algorithms for determining the equivalence of knots is extremely complex and was never implemented (Haken 1961). Hass et al. surveyed a number of knot algorithms (Hass et al. 1999). However, their conclusion was that none of them are of any practical use and that for several (general) problems in the area, it is not even clear what their exact complexity is. For instance, Hotz claimed to have developed an efficient knot-equivalence algorithm (Hotz 2008), which turned out to have a complexity of $O(2^{\frac{n}{3}})$.

Since determining the equivalence of knots is a very challenging problem in the general case, *knot invariants* have been investigated as an alternative. There are a considerable number of knot and link invariants, which are used to divide knots and links into different equivalence classes. The *multiplicity* of a link is a simple invariant for links: it is simply the number of its components. More complex invariants, such as the *unknotting number*, which counts the minimum number of times a link has to cross itself to be transformed into a trivial link, are easy to express, but hard to actually calculate. Grishanov et al. have compiled some useful invariants for classifying specific textile structures, more precisely doubly-periodic structures (Grishanov et al. 2009b).

In summary, the state-of-the-art consists of complex knot-equivalence algorithms and very specific invariants for classifying certain textile structures. None of these techniques help us in formulating a retrieval model that ranks textiles according to their similarity to a given query pattern. Furthermore, some of the methods are based on deforming and unknotting a structure, which, applied to textiles, could lead to breaking them up into individual threads. This would be contrary to what we are trying to achieve: measuring the similarity of two textile structures according to the relative positions of crossings within them. Although our work was inspired by knot theory, especially the notion of tangles, we were striving for a more practical approach. In the following section, we provide details on the inner workings of our technique.

## 4 Our Graph Textile Modelling Approach

In this section we discuss the different components of our approach: a representation of textile structures based on hypergraphs, the extraction of features (which we call neighbourhoods) from these hypergraphs, and finally a similarity measure based on neighbourhoods.

### 4.1 Textile Graphs

In a first step, we decompose a fabric into its basic building blocks, which in our case is a *crossing* of two threads together with the four links connecting it to neighbouring crossings. As already mentioned in Section 3, this is basically a 2-tangle, but we only allow a single crossing in the tangle.

**Definition 1** A *textile graph* is defined as a hypergraph $H(C, T, \Xi, \Pi, \Omega)$, where $C$ is a set of vertices that belong to crossings, $T$ a set of terminal nodes that end threads, $\Xi$ a set of hyperedges (of degree four), also called crossings, that connect vertices from $C$, $\Pi$ a set of regular edges (of degree two) that indicate which thread is on top in each crossing, and $\Omega$ a set of edges connecting vertices to vertices from other crossings or to terminal nodes.

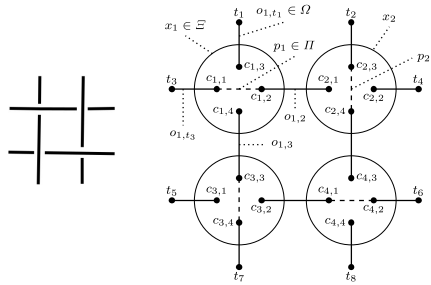

Fig. 6: A textile structure and its hypergraph $H_1$

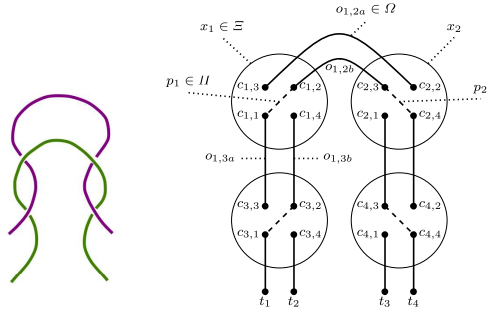

Fig. 7: Another example: hypergraph $H_2$

The hypergraphs created by the mapping of textile structures have the following properties. The cardinality of the set of vertices, $|C|$, is always a multiple of four (as these are the four endpoints of a tangle or crossing) and every $c_i \in C$ belongs to exactly one hyperedge $x_j \in \Xi$, which connects the different parts of a crossing. Associated with every $x_j$ is one (and only one) $p_j \in \Pi$, which indicates the thread on top in this crossing. Since we only allow a single crossing in a tangle, only two of the $c_i$ belonging to $x_j$ are connected via a *top edge*. Furthermore, every node $c_i$ is either connected to a node from another crossing or to a terminal node $t_i \in T$ (ending a thread). Consequently, every $c_i$ and $t_i$ show up in only one edge $o \in \Omega$ and all $t_i$ have a degree of one. The set $T$ can be empty, which means that all the threads form one or more loops, i.e., we only have knots or links. However, as we focus on knitted, woven, and similar textiles, we ignore these kinds of constructs here.

In Figures 6 and 7 we show examples of textile structures and their mapping to hypergraphs. In these figures the nodes $c_{i,j}$ belong to $C$, the $t_i$ to $T$. The solid lines $o_{i,j}$ are the edges in $\Omega$, the dashed lines $p_i$ belong to $\Pi$, while the hyperedges $x_i$ in $\Xi$ are represented by circles.

### 4.2 Comparing Textile Graphs

After defining a hypergraph representation for textiles, we now need a method to compare these graphs. Many of the approaches found in literature, such as subgraph isomorphisms for hypergraphs (Ha et al. 2018), graph editing distances for hypergraphs (Bunke et al. 2008), and morphology-based techniques (Bloch et al. 2013), have a high complexity, i.e., exponential run time.

We utilise an efficient two-phase approach for estimating the similarity of two textile graphs. In the first phase, we extract features from a textile hypergraph in the form of sets of subgraphs (we give details in Section 4.3). In the second phase, we express the similarity of two textile graphs via the similarity of sets of subgraphs (we cover this part in Section 4.5).

### 4.3 Extracting Structural Information

We were inspired by work done by Zeng et al. on traditional graphs using so-called *star structures* to represent the internal structure of graphs (Zeng et al. 2009). Essentially, a star structure is a node together with all its surrounding neighbours: a graph can then be described by a set of star structures (one for each node). Basically, a star structure is an unordered tree of depth two. Strictly speaking, the neighbours of a node in a hypergraph are also unordered, but the top edges provide some additional information we can exploit. We also go beyond the work of Zeng et al. (Zeng et al. 2009) by generating extracted subgraphs of fixed but arbitrary size. Let us start by formalising the relative positions of threads in two neighbouring crossings.

**Definition 2** Given a node $c_i$ belonging to crossing $x_i \in \Xi$ and a node $c_j$ belonging to crossing $x_j \in \Xi$ ($x_i \neq x_j$) connected by an edge $o_{i,j} = (c_i, c_j) \in \Omega$, we say that $o_{i,j}$ is

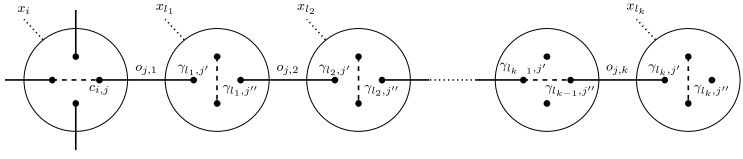

Fig. 8: Illustration of one branch of a $k$-neighbourhood

- *alternating*, if one of its endpoints is found in $\Pi$ (i.e., it is connected via a top edge) and the other is not. More formally, either $(\exists \gamma_i \in x_i : (c_i, \gamma_i) \in \Pi) \wedge (\forall \gamma_j \in x_j : (c_j, \gamma_j) \notin \Pi)$ or $(\forall \gamma_i \in x_i : (c_i, \gamma_i) \notin \Pi) \wedge (\exists \gamma_j \in x_j : (c_j, \gamma_j) \in \Pi)$.
- *non-alternating*, if both of its endpoints are either connected via a top edge or both are not. Formally, $\exists \gamma_i \in x_i, \gamma_j \in x_j : (c_i, \gamma_i) \in \Pi \wedge (c_j, \gamma_j) \in \Pi$ or $\forall \gamma_i \in x_i, \gamma_j \in x_j : (c_i, \gamma_i) \notin \Pi \wedge (c_j, \gamma_j) \notin \Pi$.
- *terminated*, if one of its endpoints is a terminator: $\exists t_j \in T : (c_i, t_j) \in \Omega$

So, an alternating thread changes positions from one crossing to the next one, either going from top to bottom or the other way around. For instance, in Figure 6 the edges $o_{1,2}$ and $o_{1,3}$ are alternating, whereas the edges $o_{1,t_1}$ and $o_{1,t_3}$ terminate. Figure 7 depicts an alternating thread from edge $o_{1,3a}$ to edge $o_{1,3b}$, while the thread from $o_{1,2a}$ to $o_{1,2b}$ is non-alternating. Also, the edge from $c_{3,1}$ to $t_1$ is terminated.

*4.3.1 Neighbourhood*

We now define the *neighbourhood* of a crossing, which is essentially a 2-tuple with two sets of labels. The first set describes the behaviour of the edges connected to the top edge vertices, i.e., it specifies whether these edges are alternating ('a'), non-alternating ('n'), or connect to a terminal ('t'). The second set describes this for the edges connected to the bottom edge vertices.

**Definition 3** Given a crossing defined by $x_i \in \Xi$, let $c_{i,1}$ and $c_{i,2}$ stand for the top thread, i.e. $(c_{i,1}, c_{i,2}) \in \Pi$ and $c_{i,3}$ and $c_{i,4}$ for the bottom thread, i.e. $(c_{i,3}, c_{i,4}) \notin \Pi$. $B(x_i) = [\{z_1, z_2\}, \{z_3, z_4\}]$ is the *neighbourhood* of crossing $x_i$ where

$$z_j = \begin{cases} \text{'a' if } (c_{i,j}, \gamma_{l,j}) \in \Omega \text{ alternating} \\ \text{'n' if } (c_{i,j}, \gamma_{l,j}) \in \Omega \text{ non-alternating} \\ \text{'t' if } (c_{i,j}, t_l) \in \Omega \text{ terminated} \end{cases}$$

and the $\gamma_{l,j}$ are nodes from the other crossings that the $c_{i,j}$ connect to, or in the case of $t_l$ it is a terminator

For example, the neighbourhood of crossing 1 in Figure 6 is described by the tuple $[\{'t', 'a'\}, \{'t', 'a'\}]$. Hence, one of the main advantages of this approach becomes evident: the representation retains all the relative spatial relationships, but at the same time is orientation invariant. Rotating the textile pattern by 90 or 180 degrees or mirroring the structure has no effect on the textile graph and its crossing neighbourhoods.

We can now represent a textile hypergraph by determining the neighbourhood of every crossing in the graph and storing all the neighbourhood tuples in a multiset.

**Definition 4** The *fingerprint* $F(H)$ of a textile graph $H(C, T, \Xi, \Pi, \Omega)$ is the multiset of the neighbourhoods of its crossings: $F(H) = \{B(x_i) | x_i \in \Xi\}$

For example, the fingerprint of the textile graph shown in Figure 6 is $F(H_1) = \{[\{'a','t'\},\{'a','t'\}],[\{'a','t'\},\{'a','t'\}], [\{'a','t'\},\{'a','t'\}],[\{'a','t'\},\{'a','t'\}]\}$, which means that all the nodes of the crossings are connected to terminals or are part of alternating edges. This makes sense, as the textile shown in Figure 6 is a plain weave, which is characteristically defined by alternating threads. The fingerprint of the textile in Figure 7, on the other hand, looks different: $F(H_2) = \{[\{'a','n'\},\{'a','n'\}],[\{'a','n'\},\{'a','n'\}], [\{'a','t'\},\{'a','t'\}],[\{'a','t'\},\{'a','t'\}]\}$.

*4.3.2 k-Neighbourhood*

Next, we generalise the concept of a neighbourhood by not just looking at immediate neighbours of a crossing, but by continuing to follow a thread farther and noting whether it alternates or not. By traversing the next $k$ neighbours of the four outgoing threads of a crossing, we create a *k-neighbourhood*. In case we encounter a terminal node, the traversal stops in that direction.

**Definition 5** Given a crossing defined by $x_i \in \Xi$, again let $c_{i,1}$ and $c_{i,2}$ stand for the top thread, i.e. $(c_{i,1}, c_{i,2}) \in \Pi$ and $c_{i,3}$ and $c_{i,4}$ for the bottom thread, i.e. $(c_{i,3}, c_{i,4}) \notin \Pi$. Furthermore, let $x_{l_1,j}, x_{l_2,j}, \ldots, x_{l_k,j}$ be the sequence of $k$ crossings we encounter when following the thread leaving $x_i$ via $c_{i,j}$ and let $\gamma_{l_h,j'}, \gamma_{l_h,j''} \in x_{l_h,j}$ be the nodes in each crossing along this thread, i.e., either $\gamma_{l_h,j'}$,
$\gamma_{l_h,j''} \in \Pi$ or $\gamma_{l_h,j'}, \gamma_{l_h,j''} \notin \Pi$. The edges $o_{j,h} \in \Omega$ connect nodes from different crossings, so $o_{j,1}$ connects $c_{i,j}$ and $\gamma_{l_1,j'}$, and for $h \geq 2$, $o_{j,h}$ connects $\gamma_{l_{h-1},j''}$ and $\gamma_{l_h,j'}$. Figure 8 illustrates this situation. Then $B_k(x_i) = [\{y_1, y_2\}, \{y_3, y_4\}]$ is the *k-neighbourhood* of crossing $x_i$ with $y_j = [y_{j,1}, y_{j,2}, \ldots, y_{j,k}]$ where each

$$y_{j,m} = \begin{cases} 'a' & \text{if } o_{j,m} \in \Omega \text{ alternating} \\ 'n' & \text{if } o_{j,m} \in \Omega \text{ non-alternating} \\ 't' & \text{if } o_{j,m} \in \Omega \text{ terminated} \end{cases}$$

If $o_{j,m}$ is a terminated edge for $m < k$, we only have $m$ elements in tuple $y_j$. For the example in Figure 8 the tuple $y_j$ is equal to $['a','n',\ldots,'a']$.

This makes the neighbourhood described in Definition 3 a special case of a $k$-neighbourhood with $k = 1$. The 2-neighbourhood of crossing $x_1$ in Figure 7, for example, is $[\{['a','t'],['n','a']\}, \{['n','a'],['a','t']\}]$. The fingerprints of hypergraphs using $k$-neighbourhoods are computed accordingly, we just have to replace $B(x_i)$ with $B_k(x_i)$ in Definition 4.

4.4 Implementation

We now turn to implementation issues. Algorithm 1 shows how to compute the fingerprint of a hypergraph in pseudo-code. We go through all the crossings of a hypergraph and follow the outgoing threads from the four nodes of this crossing to the next $k$ crossings to explore its neighbourhood. W.l.o.g., we call the nodes connected via the top edge $c_{i,1}$ and $c_{i,2}$, i.e., $(c_{i,1}, c_{i,2}) \in \Pi$, and we denote the nodes of the bottom thread $c_{i,3}$ and $c_{i,4}$, i.e., $(c_{i,3}, c_{i,4}) \notin \Pi$. For a node in a crossing, the function NEIGHBOUR gives us the node in a neighbouring crossing that it is connected to. The function FINDLABEL returns the label of an edge and, given a node, the function OPPOSITE gives us the node located on the other side of a crossing. If we encounter a terminal node before visiting $k$ neighbouring crossings, we pad the labels with NULL values.

**Algorithm 1:** FINGERPRINT($H$,$k$)

**Input** : hypergraph $H(C, T, \Xi, \Pi, \Omega)$ with
      $C$: set of vertices belonging to crossings
      $T$: set of terminals
      $\Xi$: set of hyperedges connecting vertices
      $\Pi$: set of edges indicating top thread
      $\Omega$: set of edges connecting crossings/terminals
      $k$: size of the neighbourhood
**Output:** fingerprint for hypergraph $H$

1   FP := ∅;
2   **for** *every $x_i \in \Xi$* **do**
3      **for** *$j := 1$ to $4$* **do**
4         let $c_{i,j}$ be the $j$-th node of crossing $x_i$;
5         (* $j = 1, 2$ for top-level thread *)
6         $l := 0$, current $= c_{i,j}$;
7         **repeat**
8            $l$++;
9            $\gamma_{l,j'}$ := NEIGHBOUR(current);
10           label$_{j,l}$ := FINDLABEL(current, $\gamma_{l,j'}$);
11           **if** *label$_{j,l} \neq$ 't'* **then**
12              $\gamma_{l,j''}$ := OPPOSITE($\gamma_{l,j'}$);
13              current := $\gamma_{l,j''}$;
14           **end**
15         **until** *$l = k$ or label$_{j,l}$ = 't'*;
16         **for** *$s := l + 1$ to $k$* **do**
17            label$_{j,s}$ := NULL;
18         **end**
19      **end**
20      fp$_i$ := [{[label$_{1,1}$, label$_{1,2}$, ..., label$_{1,k}$], [label$_{2,1}$, label$_{2,2}$, ..., label$_{2,k}$]},
21            {[label$_{3,1}$, label$_{3,2}$, ..., label$_{3,k}$], [label$_{4,1}$, label$_{4,2}$, ..., label$_{4,k}$]}];
22      FP := FP ∪ fp$_i$;
23   **end**
24   return FP;

In order to implement our algorithm efficiently, we use the following data structure to store nodes of a crossing in a hypergraph:

```
struct node {
  int   nextNode;
  bool  onTop;
  int   oppositeNode;
}
```

We store all the nodes of a hypergraph in an array, taking care to place the four nodes of a crossing into four consecutive cells of the array. So, the nodes at positions $4i$ to $4i + 3$ belong to crossing $i$ (for $0 \leq i \leq n - 1$, assuming we have $n$ crossings). `nextNode` contains the index of the node of the neighbouring crossing that the current `node` connects to. We set this value to -1 if it connects to a terminal node.[2] The Boolean `onTop` tells us whether `node` belongs to the top edge of a crossing or not. Strictly speaking, we do not need `oppositeNode`: we could check all the nodes belonging to the same crossing as `node` and find the node with the same value for `onTop`. However, doing so would be very inefficient.

For the overall complexity of the algorithm, this means that we can compute the $k$-neighbourhoods of all $n$ crossings of a textile hypergraph in $O(nk)$.

4.5 Similarity Measures

Having defined fingerprints of textile patterns for the first phase, for the second phase we now have to specify how to actually measure their similarity or distance. The Euclidean distance, the cosine measure, the Hamming distance, the Jaccard distance and the overlap coefficient are all typical distance metrics employed for measuring similarity (Leskovec et al. 2014; Zaki and Meira 2014), which we also apply to our fingerprints.

However, the fingerprints of our textile graphs are multisets rather than sets, i.e., for every element in a multiset we store the number of occurrences of the element. For instance, the multiset $\{a, a, a, b, c, c\}$ becomes $\{a : 3, b : 1, c : 2\}$ or just $(3, 1, 2)$ if we assign fixed positions to each element. In this case, position 1 represents the frequency of $a$, position 2 the frequency of $b$, and position 3 the frequency of $c$. Fixing the positions of the elements within the multisets allows us to interpret them as points or vectors.

Using the point or vector representation of two two multisets $R = (r_1, r_2, \ldots, r_n)$ and $S = (s_1, s_2, \ldots, s_n)$ allows us to apply the Euclidean distance, the cosine measure, the Hamming distance, the Jaccard distance, and the overlap coefficient for measuring the distance between $R$ and $S$. We go into more details in the following.

4.5.1 Euclidean Distance

We can calculate the Euclidean distance between two points (using a vector representation):

$$D_E(R, S) = \sqrt{\sum_{i=1}^{n} (r_i - s_i)^2} \quad (1)$$

Computing $D_E(F(H_1), F(H_2))$ gives us the distance between two textile graphs $H_1$ and $H_2$. For example, the fingerprint of $H_1$ (Figure 6), using 1-neighbourhood, is made up of four times the tuple [{'a','t'},{'a','t'}] and does not contain [{'a','n'},{'a','n'}], while the fingerprint of $H_2$, using 1-neighbourhood, contains both of these tuples two times each, i.e., we can represent $H_1$ by $(0,4)$ and $H_2$ by $(2,2)$. Applying Formula (1), we obtain $\sqrt{(0-2)^2+(4-2)^2}=2\sqrt{2}$.

---

[2] We do not store the terminal nodes explicitly.

*4.5.2 Cosine Measure*

We can also apply the cosine distance, which applies an inner product, to a vector representation:

$$D_c(R,S) = 1 - \frac{\sum_{i=1}^{n} r_i \cdot s_i}{\sqrt{\sum_{i=1}^{n} r_i^2} \cdot \sqrt{\sum_{i=1}^{n} s_i^2}} \quad (2)$$

$D_c(F(H_1), F(H_2))$ computes the cosine measure distance between two textile hypergraphs $H_1$ and $H_2$. For instance, using Formula (2) on the frequency vectors of the fingerprints (using a 1-neighbourhood) of $H_1 = (0,4)^\mathsf{T}$ and $H_2 = (2,2)^\mathsf{T}$ yields $1 - {}^{0\,+\,8}\!/\!{4\sqrt{8}} = 1 - \sqrt{2}/2$.

The law of diminishing returns also applies to term frequencies within individual documents and in the whole document collection. The more frequently a term appears in a document, the smaller the additional impact will be. Also, terms appearing less frequently in document collections tend to be more important. Term frequency (TF) and inverse document frequency (IDF) factors are used to counter this effect. These factors are applied to the input vectors, we use logarithmic TF-IDF factors: $\text{TF}_{p,t} = 1 + \log(f_{p,t})$ and $\text{IDF}_p = \log \frac{N}{f_p}$ where $f_{p,t}$ is the frequency of fingerprint $p$ in textile $t$, $N$ is the overall number of textiles in the collection, and $f_p$ is the number of textiles in the collection in which fingerprint $p$ occurs.

*4.5.3 Hamming Distance*

Mathematically, the Hamming distance counts the number of components that are different in two vectors. Let $f_H(r_i, s_i)$ be a function comparing the components $r_i$ and $s_i$, then:

$$f_H(r_i, s_i) = \delta_{r_i, s_i} = \begin{cases} 0 \text{ if } r_i = s_i \\ 1 \text{ if } r_i \neq s_i \end{cases}$$

A formal definition of the Hamming distance is equal to:

$$D_H(R,S) = \sum_{i=1}^{n} f_H(r_i, s_i) \quad (3)$$

For the same example, applying the hamming distance of the frequency vectors of the fingerprints, using 1-neighbourhood, of $H_1 = (0,4)^\mathsf{T}$ and $H_2 = (2,2)^\mathsf{T}$ return $1+1=2$.

The basic Hamming distance suffers from a few issues. There are a few issues with the Hamming distance. First, there is a lack of normalisation, i.e., the distance between two vectors can range anywhere from 0 to $n$, which even varies depending on the size of the vectors. Second, if there is a total order on the elements of the domain, then users have an intuition on how close or distant these elements should be. For example, with integer vectors, intuitively $(1,0,3)^\mathsf{T}$ is closer to $(1,0,2)^\mathsf{T}$ than $(1,0,7)^\mathsf{T}$. The basic Hamming distance would return a distance of 1 for both cases, though. This can be fixed by redefining the Hamming distance: $f_{\tilde{H}}(r_i, s_i) = |r_i - s_i|$. Consequently, the distance measure becomes

$$D_{\tilde{H}}(R, S) = \sum_{i=1}^{n} f_{\tilde{H}}(r_i, s_i) = \sum_{i=1}^{n} |r_i - s_i| \qquad (4)$$

Applying $H_1 = (0,4)^\mathsf{T}$ and $H_2 = (2,2)^\mathsf{T}$ to Formula (4) gives us $2 + 2 = 4$.

### 4.5.4 Jaccard Coefficient

The Jaccard coefficient is one of the most common similarity measures for (multi-)sets. Given the multisets $R$ and $S$, $|R \cap S|$ is computed as $\sum_{i=1}^{n} \min(r_i, s_i)$ and $|R \cup S| = \sum_{i=1}^{n} \max(r_i, s_i)$. Putting this together yields the Jaccard coefficient distance:

$$D_J(R, S) = 1 - \frac{\sum_{i=1}^{n} \min(r_i, s_i)}{\sum_{i=1}^{n} \max(r_i, s_i)} = \frac{\sum_{i=1}^{n} |r_i - s_i|}{\sum_{i=1}^{n} \max(r_i, s_i)} \qquad (5)$$

Applying Formula (5) to the frequency vectors of the 1-neighbourhood fingerprints of $H_1 = (0,4)^\mathsf{T}$ and $H_2 = (2,2)^\mathsf{T}$ gives us $2 + 2/2 + 4 = 2/3$.

### 4.5.5 Overlap Coefficient

The overlap coefficient, which is related to the Jaccard coefficient, takes the cardinality of the intersection of the two sets and divides it by the cardinality of the smaller of the two sets. So, with $|R \cap S| = \sum_{i=1}^{n} \min(r_i, s_i)$ and $\min(|R|, |S|) = \min(\sum_{i=1}^{n} r_i, \sum_{i=1}^{n} s_i)$, the overlap coefficient distance measure between multisets $R$ and $S$ is

$$D_o(R, S) = 1 - \frac{\sum_{i=1}^{n} \min(r_i, s_i)}{\min(\sum_{i=1}^{n} r_i, \sum_{i=1}^{n} s_i)} \qquad (6)$$

For the same example; $H_1 = (0,4)^\mathsf{T}$ and $H_2 = (2,2)^\mathsf{T}$, the overlap coefficient is: $D_o(H_1, H_2) = 1 - 0 + 2/min(4,4) = 1 - 2/4 = 1/2$.

## 5 Textile Retrieval and Clustering

When developing our hypergraph-based approach, we had two applications in mind. On the one hand, we wanted to apply it in the ranked retrieval of textile structures from a collection. On the other hand, we wanted to see whether our method is suitable for unsupervised learning techniques such as clustering. The textile retrieval and clustering techniques help to find and determine the structure or kind of the textile patterns. From that, we can know their particular applications, materials, fabrication methods and origins. The techniques allow domain experts to gain deeper insights about quantify differences and variations of textiles in different time and cultures. They also support some algorithms for detecting density and defect of a textile. Specially, in the paper, the retrieval and clustering are used to show the accuracy or performance of the similarity measure.

5.1 Retrieval

An algorithm for ranked retrieval is quite straightforward (see Algorithm 2). We just have to compute the similarity between each textile in a collection and a query and then sort the result by this similarity. The crucial part of the algorithm is the distance measure $d_m$ being $D_E$, $D_c$, $D_H$, $D_{\tilde{H}}$, $D_J$ or $D_o$ described in Section 4.5. In Section 7 we evaluate the performance of the different measures for ranked retrieval.

**Algorithm 2:** RANK($S$, $q$, $d_m$)

**Input** : $S$: set of $n$ hypergraphs $\{H_1, H_2, \ldots, H_n\}$
 $q$: query pattern
 $d_m$: distance measure between hypergraphs
**Output:** $R$: ranked list of $S$ according to the similarity to $q$

1 $R := \emptyset$;
2 **for** $i := 1$ *to* $n$ **do**
3 $\quad r_i := d_m(F(H_i), F(q))$;
4 $\quad R := R \cup \{(H_i, r_i)\}$;
5 **end**
6 order tuples in $R$ by descending $r_i$;
7 return $R$;

5.2 Clustering

Basically, clustering is a classification task grouping $n$ textile hypergraphs into $m$ clusters of textiles with similar weaving structures. For the cluster algorithms we used the well-known methods of hierarchical agglomerative clustering (HAC) and K-means (Yildirim et al. 2018), using our textile modelling approach to measure distances between textiles. In the following, we describe HAC and K-means in more detail.

5.3 Hierarchical Agglomerative Clustering (HAC)

Algorithm 3 shows a basic version of hierarchical agglomerative clustering in pseudo-code. Each textile hypergraph is treated as a single cluster at initiation, and then pairs of clusters are merged (or agglomerated) as we move up the hierarchy until, finally, we have $m$ clusters in the active set $L$. When two clusters are merged, they are removed from $L$ and their union is added to $L$.

The algorithm has a time complexity $O(n^2 d \log n)$ to find $m$ clusters from $n$ patterns having $d$ dimensions.

In the algorithm, we use the function DISTANCEMATRIX to compute a distance matrix $\chi$ containing all the pairwise distances between the fingerprints of all the hypergraphs in $S$. For that purpose, we use the distance measure $d_m$, which is one of the measures described in Section 4.5. The function TWOCLOSEST determines the two clusters $u_1$ and $u_2$ in $L$ that are closest to each other. For this, we need the distance matrix $\chi$ and a criterion $d_c$, which defines how to compute the distance between two sets of hypergraphs. Commonly used criteria in HAC are Ward's method, single-linkage, complete-linkage, and average-linkage, which are described in the following.

**Algorithm 3:** HAC($S$, $m$, $d_m$, $d_c$)

**Input** : $S$: set of $n$ hypergraphs $\{H_1, H_2, \ldots, H_n\}$
$m$: number of clusters
$d_m$: distance measure between hypergraphs
$d_c$: distance criterion between clusters
**Output:** $L$: set of $m$ clusters

1  $\chi :=$ DISTANCEMATRIX($S, d_m$);
2  $L := \emptyset$;
3  **for** $i := 1$ *to* $n$ **do**
4  $\quad$ $L := L \cup \{\{H_i\}\}$;
5  **end**
6  **while** $|L| > m$ **do**
7  $\quad$ $u_1, u_2 :=$ TWOCLOSEST($L, \chi, d_c$);
8  $\quad$ $L := (L \setminus u_1) \setminus u_2$;
9  $\quad$ $L := L \cup \{u_1 \cup u_2\}$;
10 **end**
11 **return** $L$;

Ward's criterion considers the squared (Euclidean) distance between the centroids of two clusters:

$$DC_W(u_i, u_j) = \frac{|u_i||u_j|}{|u_i| + |u_j|} D_E(F(c_i), F(c_j))^2 \qquad (7)$$

where $c_l$ is the centroid of cluster $u_l$ and is defined as:

$$c_l = \frac{1}{|u_l|} \sum_{H_i \in u_l} F(H_i) \qquad (8)$$

The single-linkage and complete-linkage criteria look at the minimal and maximal distance, respectively, over all possible hypergraph pairs from different clusters:

$$DC_S(u_i, u_j) = \min_{H_r \in u_i, H_s \in u_j} d_m(F(H_r), F(H_s)) \qquad (9)$$

$$DC_C(u_i, u_j) = \max_{H_r \in u_i, H_s \in u_j} d_m(F(H_r), F(H_s)) \qquad (10)$$

Rather than just considering the minimal and maximal distance, the average-linkage criterion averages over all possible hypergraph pairs between the clusters:

$$DC_A(u_i, u_j) = \frac{1}{|u_i||u_j|} \sum_{H_r \in u_i} \sum_{H_s \in u_j} d_m(F(H_r), F(H_s)) \qquad (11)$$

5.4 K-Means

Algorithm 4 depicts the pseudocode for the K-means algorithm, which finds the clusters based on centroids.[3] Initially, $m$ centroids are picked randomly, we call the (current) set of clusters $\mathtt{C}$. There are different strategies for picking this initial set, but common to these strategies is to put them near the data points and well apart from each other. We use the function RANDOM($S$, $m$) that randomly selects $m$ items from $S$. Every centroid defines a cluster and we assign every textile pattern $H_i$ to the centroid closest to it. The function CLOSEST($\mathtt{C}$, $H_i$, $d_m$) finds the centroid in $\mathtt{C}$ closest to $H_i$, according to distance measure $d_m$. After assigning all hypergraphs, we move each centroid to the average or mean location of the data points assigned to it by recomputing it using Formula (8). We repeat the assignment and recomputation step until there is no (or very little) change. As there is no guarantee that the algorithm will converge, we also define a number of maximum iterations after which the algorithm stops. In terms of complexity, finding an optimal configuration that minimises the overall distances of all data points to their respective centroids is NP-hard. This is another reason to run a version of the algorithm with a parameter $max$ for the maximum number of iterations. When limiting the number of iterations, the complexity of the algorithm is $O(max \cdot m \cdot n \cdot d)$, where $d$ is the dimension of the data points.

**Algorithm 4:** K-MEANS($S$, $m$, $max$, $d_m$)

**Input** : $S$: set of $n$ hypergraphs $\{H_1, H_2, \ldots, H_n\}$
  $m$: number of clusters
  $max$: maximum number of iteration
  $d_m$: distance measure between hypergraphs
**Output:** $\mathtt{C}$: set of $m$ centroids $\{c_1, c_2, \ldots, c_m\}$
  $L$: set of $m$ clusters $\{u_1, u_2, \ldots, u_m\}$

1  $\mathtt{C} := \text{RANDOM}(S, m)$;
2  $iter := 0$;
3  $change := \text{true}$;
4  **while** $(iter \leq max) \wedge change$ **do**
5  $\quad iter++$;
6  $\quad$ **for** $k := 1$ *to* $m$ **do**
7  $\quad\quad u_k := \emptyset$;
8  $\quad$ **end**
9  $\quad$ **for** $i := 1$ *to* $n$ **do**
10 $\quad\quad j := \text{CLOSEST}(\mathtt{C}, H_i, d_m)$;
11 $\quad\quad u_j := u_j \cup \{H_i\}$;
12 $\quad$ **end**
13 $\quad change := \text{false}$;
14 $\quad$ **for** $k := 1$ *to* $m$ **do**
15 $\quad\quad$ compute new centroid $c_k^{new}$ of $u_k$
16 $\quad\quad$ **if** $c_k \neq c_k^{new}$ **then**
17 $\quad\quad\quad change := \text{true}$;
18 $\quad\quad\quad c_k := c_k^{new}$
19 $\quad\quad$ **end**
20 $\quad$ **end**
21 **end**
22 **return** $\mathtt{C}$, $L$;

---

[3] Usually, $k$ stands for the number of clusters, which we already use for neighbourhoods. Thus, we use the parameter $m$ for the number of clusters.

## 6 Evaluation Methodology

We evaluated the different variants of our similarity measure experimentally, clustering a data set containing $1,600$ textiles and comparing the outcome to the correct classification. Additionally, we run queries on our data set, measuring the retrieval performance. We also look at the linear complexity, presenting numbers on the execution time of the algorithm.

### 6.1 Experiment Setup

The algorithms were implemented using Java 1.8.0_171 running under Windows 10 (64-bit). All experiments were run on a computer with an Intel Core i7 CPU (2.40 GHz) and 16 GB memory.

We evaluate the retrieval performance by using each of the textile objects as a query $q_i \in Q$ and then ranking all the other textiles according to their similarity to the query. All $m_i$ textiles $\{h_1, h_2, \ldots, h_{m_i}\}$ that are in the same category $\alpha_i$ as $q_i$ are considered to be relevant, while those from other categories are not relevant.

Testing the effectiveness of our similarity measure for clustering boils down to the following. We use our approach to divide up a collection of $n$ textile hypergraphs $S = \{H_1, H_2, \ldots, H_n\}$ into $m$ clusters $L = \{\lambda_1, \lambda_2, \ldots, \lambda_m\}$ and then compare the result to the correct classification $A = \{\alpha_1, \alpha_2, \ldots, \alpha_m\}$.

### 6.2 Quality Measures

We now take a closer look at how we measure the quality of the clustering and retrieval performance.

#### 6.2.1 Retrieval Performance

We measure the quality of the resulting ranked lists using mean average precision (MAP), mean precision at 100 (MeanP@100), average Precision-Recall (PR), and average F-measure-Recall (FR) (Manning et al. 2009). Essentially, MAP aggregates the quality across all recall levels into a single number:

$$MAP(Q) = \frac{1}{|Q|} \sum_{i=1}^{|Q|} \frac{1}{m_i} \sum_{j=1}^{m_i} Precision(R_{ij})$$

where $R_{ij}$ is a ranked list (from the first textile down to $h_j$ which belongs to the returned relevant textiles $\{h_1, h_2, \ldots, h_{m_i}\}$ of the query $q_i \in Q$) and $Precision(R_{ij})$ is the precision of $R_{ij}$.

In many cases, users are not interested in going through all the returned results, which makes the precision at $k$ documents (P@k) a useful metric. In our data set, each query has 100 related textiles, so we use P@100 to evaluate the precision of each query $q_i \in Q$ for the first 100 results. We compute the mean precision MeanP@100 for all queries in $Q$ as follows:

$$MeanP@100(Q) = \frac{1}{|Q|} \sum_{i=1}^{|Q|} P@100(q_i)$$

A PR curve plots the precision against the recall; we use the the standard 11-point interpolated average precision here. The interpolated precision of query $q_i$ at the standard recall level $r_l$, $0 \leq l \leq 10$, is defined as the highest precision found for any recall level $r \geq r_l$: $P_i(r_l) = \max_{r \geq r_l} P_i(r)$, where $P(r)$ is the precision at recall level $r$. Thus, the average precision of $Q$ at the standard recall level $r_l$ is equal to:

$$\bar{P}(r_l) = \frac{\sum_{i=1}^{|Q|} P_i(r_l)}{|Q|}$$

Similarly, the average F-measure of $Q$ at the standard recall level $r_l$ is defined as:

$$\bar{F}(r_l) = \frac{\sum_{i=1}^{|Q|} F_i(r_l)}{|Q|}$$

where $F_i(r_l) = \frac{2P(r_l)r_l}{P(r_l)+r_l}$ is the interpolated F-measure of query $q_i$ at $r_l$.

### 6.2.2 Clustering

For the purpose of measuring the quality of the clustering $L$ compared to the correct classification $A$, we apply purity, normalised mutual information (NMI), Rand index, precision, recall, and F-measure (Manning et al. 2009).

Purity is a simple and transparent evaluation measure counting the number of correctly classified textiles. To do so, we assign each cluster $\lambda_i \in L$ to the class $\alpha_j \in A$ that has the largest overlap with $\lambda_i$ and count the number of shared elements. We normalise the result by dividing by the total number of textiles:

$$purity = \frac{1}{n} \sum_{\lambda_i \in L} \max_{\alpha_j \in A} |\lambda_i \cap \alpha_j|$$

The larger the number of clusters, the easier it is to achieve high purity. In the extreme case of creating $n$ clusters (one for each textile), we would achieve a purity of 1. This makes it difficult to compare the quality of clusterings that have a different number of clusters.

This has led to the utilisation of normalised mutual information (NMI), which is based on concepts from information theory, such as entropy. Given two random variables, the mutual information tells us how the uncertainty of one of them decreases by being aware of the other one. For clustering this means: how much knowledge do we gain about the classification $A$ knowing the clustering $L$? NMI is defined as:

$$NMI = \frac{I(L;A)}{(H(L)+H(A))/2}$$

where $I(L;A)$ is the mutual information shared by $L$ and $A$:

$$I(L;A) = \sum_{\lambda_i \in L} \sum_{\alpha_j \in A} \frac{|\lambda_i \cap \alpha_j|}{n} \log_2 \frac{n|\lambda_i \cap \alpha_j|}{|\lambda_i||\alpha_j|}$$

and $H(L)$ and $H(A)$ measure the entropy of $L$ and $A$, respectively:

$$H(L) = -\sum_{\lambda_i \in L} \frac{|\lambda_i|}{n} \log_2 \frac{|\lambda_i|}{n}$$

$$H(A) = -\sum_{\alpha_j \in A} \frac{|\alpha_j|}{n} \log_2 \frac{|\alpha_j|}{n}$$

Last, but not least, the Rand index categorises every pair of hypergraphs $H_i, H_j \in S, i > j$ as either a true positive (TP), a true negative (TN), a false positive (FP), or a false negative (FN). The categorisation depends on which condition a pair satisfies:

- TP: $H_i$ and $H_j$ are in the same cluster in $L$ and in the same class in $A$.
- TN: $H_i$ and $H_j$ are in different clusters in $L$ and in different classes in $A$.
- FP: $H_i$ and $H_j$ are in the same cluster in $L$ and in different classes in $A$.
- FN: $H_i$ and $H_j$ are in different clusters in $L$ and in the same class in $A$.

Basically, the Rand index (RI) determines the ratio of textiles placed into the correct cluster:

$$RI = \frac{TP+TN}{TP+FP+TN+FN}$$

The Rand index puts the same emphasis on all these factors. However, when categorising pairs, there is usually a bias: it is much easier to identify true negatives correctly, due to their large number. That is why we also look at the standard quality measures of precision $P = \frac{TP}{TP+FP}$, recall $R = \frac{TP}{TP+FN}$, and F-measure $F = \frac{2PR}{P+R}$.

6.3 Data Set

In an earlier project, we developed a textile editor called SAWU that allows a user to enter textile patterns via a graphical user interface by creating thread crossings and connecting the ends of the crossings with each other. Additionally, a user can cut, copy, and paste subpatterns and reconnect them to other parts of a textile. Complex and irregular patterns have to be constructed manually, while simple recurring patterns can be automatically generated and then modified if need be. More details on the editor can be found in (Martins et al. 2013; Győry 2014). Essentially, a user can construct large complex textile patterns from simple building blocks. We use the output of the textile editor, consisting of text files, to generate the corresponding hypergraphs.

With the help of domain experts we selected sixteen important categories of textiles, each with 100 specimens, resulting in a data set containing a total of 1600 fabrics. This data set is used to evaluate the clustering performance over the sixteen categories and the retrieval performance by utilising each of the 1600 fabrics as a query. We uploaded and shared this data set in Harvard Dataverse[4]. On average, each textile consists of 20,916 vertices, 5,229 hyperedges, 152 terminal nodes, 5,229 regular edges and 10,534 connected edges. Each textile is represented by a fingerprint based on the $k$-neighbourhood of the crossings in its hypergraph. Figure 9 gives an overview of the number of different $k$-neighbourhoods found in the data set for different values of $k$. It also shows the average number of different $k$-neighbourhoods per textile. Clearly, increasing the value for $k$ leads to a considerable increase in the number of different patterns found in a textile. While raising the value for $k$ results in slower processing speed, it helps in distinguishing textiles more accurately. Later on, we show how to balance the trade-off between speed and accuracy.

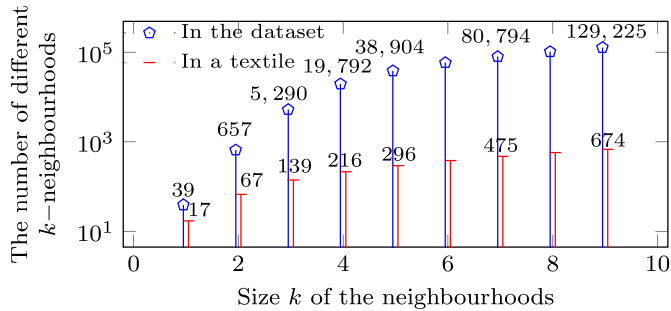

Fig. 9: Number of patterns for different values of $k$

In the following, we give an overview of the different kinds of textiles found in each group. One of the simplest weaving patterns is plain weave, in which a weft thread alternates between going over and under a warp thread.[5] In each row, this pattern is shifted by one position (see Figure 10(a)). The next five groups of patterns consist of twills, in which more than one warp thread is crossed over or under. Figures 10(b) to (f) show example patterns, ranging from 2/1 twill to 4/4 twill. In the satin (also known as sateen) weave structure (see Figure 10(g)), four or even more weft threads float over a warp thread or vice-versa. The most complex patterns in our collection are taken from a collection of weavings originating in the Andes (South America) and Vietnam (Southeast Asia). Since they were created manually, they can exhibit a great variety of different styles in a single textile. The Andean pattern depicted in Figure 10(h) and the Vietnamese weaving pattern, describing elephants, depicted in Figure 10(i) indicate this, as the warp and weft threads cross a different number of threads in different parts of the textile. For more examples, please see `http://www.weavingcommunities.org/` for Andean weavings and Figure 18 in Appendix 10.1 for Vietnamese weavings.

---

[4] https://dataverse.harvard.edu/dataset.xhtml?persistentId=doi:10.7910/DVN/ZFNLES
[5] Warp threads are longitudinal threads held in place by a frame, while the weft thread is led through the warp threads.

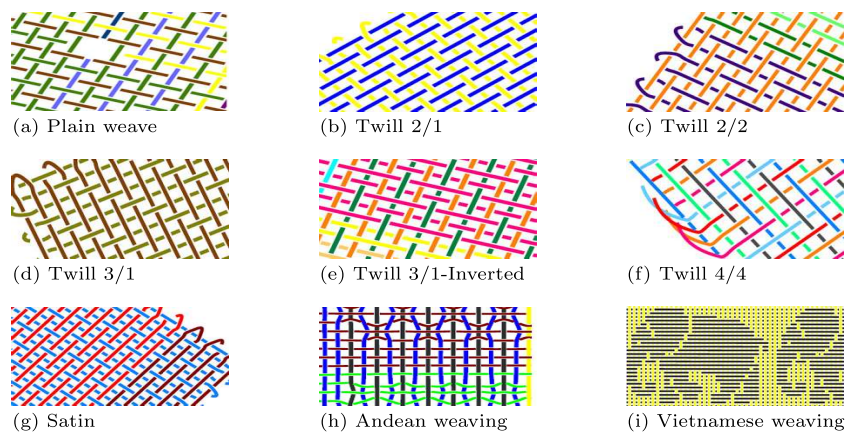

Fig. 10: Examples of weaving patterns

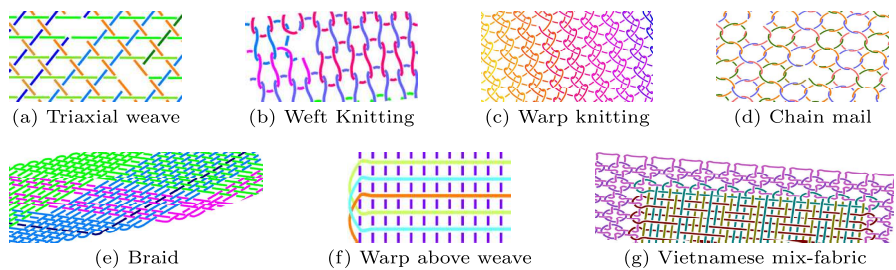

Fig. 11: More examples of textile patterns

For the remaining groups of textiles, shown in Figure 11 we have chosen patterns that are not actually woven to see how our textile recognition would cope with non-weaving patterns. Triaxial weave, although called a weave, is a hybrid structure between weaving and braiding. The resulting structure, an example of which can be seen in Figure 11(a), does not follow a rectilinear pattern. In knitting, multiple loops of yarn in a line or tube are formed by connecting a row of new loops to a row of already existing loops. When done manually, this usually involves needles holding the thread. The two basic varieties of knitting are weft knitting (see Figure 11(b)) and warp knitting (see Figure 11(c)). In weft knitting, the more common technique, the wales[6] are perpendicular to the course of the yarn and the fabric can be produced from a single yarn. By contrast, in warp knitting, the wales run roughly parallel and one yarn is required for every wale. Chain mail, shown in Figure 11(d)), is made of small rings linked together to form a mesh, which can slide against each other to create a flexible fabric. Braids are created by intertwining three or more threads as shown in Figure 11(e). In the warp above weave pattern all the threads of one type are always located above the other (see Figure 11(f)). For the most complex non-weaving patterns we have chosen Vietnamese mix-fabrics, which are hybrid structures combining two or more types of techniques, such as chain mail, braiding, and knitting. For an example, see the textile shown in Figure 11(g), in which chain mail is combined with complex weaving to define a window. Further examples of Vietnamese mix-fabric patterns can be found in Figure 19 in Appendix 10.1.

---

[6] A wale is a column of loops produced by the same needle.

We have also introduced imperfections into some of the textiles in each group to test the similarity measure's capability to deal with errors in a pattern. Additionally, we also rotated and mirrored some of the textile samples to check that our similarity measure can cope with differently oriented versions of the same weaving pattern. For that purpose, we randomly modified 1% of the crossings in the data set; 85% of the patterns were rotated in some way and 35% mirrored (this adds up to more than 100%, because textile patterns can be rotated and mirrored).

# 7 Experimental Results

Basically, two parameters are crucial for the calibration of our model: the size k of the neighbourhoods and the distance metric used for comparing two fingerprints (the Euclidean, frequency cosine, TF-IDF cosine, Boolean Hamming, frequency Hamming, Jaccard and Overlap distances; see also Section 4.5). In the following we investigate the impact of both parameters on the execution time and on the retrieval and cluster performance of our algorithms. Additionally, for the cluster performance, we investigate the impact of the unsupervised learning model (hierarchical agglomerative clustering and K-means; see also Section 5).

## 7.1 Execution Time

Figure 12 illustrates how the execution time varies with increasing neighbourhood sizes for the different distance metrics. Every data point in Figure 12 averages the execution time of nine runs each generating a complete distance matrix including the results for the pairwise comparisons of all textiles. In general, the execution time of each variant of our algorithm increases linearly with the neighbourhood size $k$, which is a highly desirable property, as it leads to a scalable solution.

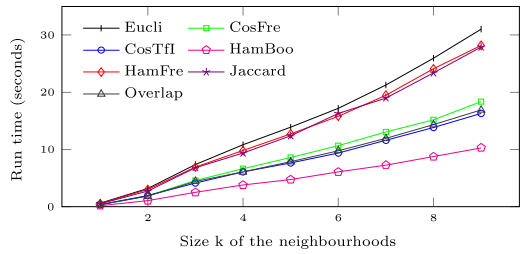

Fig. 12: Execution Time

When implementing the multisets, we refrained from using an explicit vector representation because of the sparsity of the vectors. For instance (as shown in Figure 9), although there are 129,225 different (potential) neighbourhoods for $k = 9$, on average only 674 appear in a given textile structure. As a consequence, we only need to store and look up the values not equal to zero. In our case we implemented the vectors using a HashMap.

Unsurprisingly, the Boolean Hamming distance (HamBoo), being the simplest formula, is fastest. The other distance measures are divided into two groups. Both cosine (CosFre and CosTfI) and the Overlap measures are easier to compute, as we only need to consider non-zero entries for both vectors and some additional computations for the normalisation. For the Euclidean (Eucli) and the frequency Hamming (HamFre) distances, the calculation of the differences between vector components takes more effort, while for Jaccard the normalisation is more costly.

### 7.2 Retrieval Performance

Figure 13(a) depicts the mean average precision (MAP) of the different techniques and indicates the overall utility of our similarity measure. There is no significant gain in using neighbourhoods with a size greater than four. The Jaccard, frequency cosine (CosFre), TF-IDF cosine (CosTfI), and Overlap variants are clearly on top (except TF-IDF cosine and Overlap for $k$ equal to one), with Jaccard being slightly better than CosFre, CosTfI and Overlap. At the other end, the Boolean Hamming (HamBoo) distance is always lagging behind. In the middle group, Euclidean (Eucli) and frequency Hamming (HamFre) are roughly comparable and trade places at $k$ equal to three. For $k = 4$, sorting the measures in decreasing order of precision yields: Jaccard (0.91), frequency cosine (0.897), TF-IDF cosine (0.889), Overlap (0.887), Euclidean (0.814), frequency Hamming (0.805), and Boolean Hamming (0.661).

Figure 13(b) displays the mean precision for the first 100 retrieved textile patterns (MeanP@100). The results are very similar to the ones for MAP. Jaccard is leading the pack with frequency cosine, Overlap and TF-IDF cosine being not far behind. The other three distance measures are worse and show a very similar relative positioning as for MAP. Euclidean and frequency Hamming are very close to each other and trade places for $k = 4$. For $k = 4$, sorting the measures in decreasing order of precision yields: Jaccard (0.881), frequency cosine (0.864), Overlap (0.863), TF-IDF cosine (0.855), frequency Hamming (0.777), Euclidean (0.772), and Boolean Hamming (0.597).

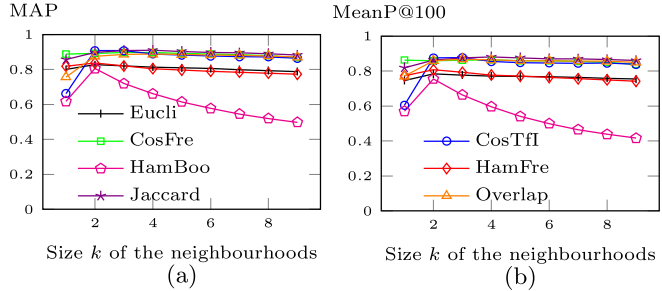

Fig. 13: Mean Average Precision and Mean Precision at 100

Figure 14 shows the average Precision Recall (PR) and average F-measure Recall (FR) curves for neighbourhoods of size four. Again, the Jaccard distance shows excellent results, being on top in the PR and FR curves. In the PR curve, its precision stays above 90% for recall values up to 60%, above or equal 85% for recall values from 70% to 90% and then drops to around 68%. In the FR curve, it achieves a maximum F-measure of 87.2 for a recall value of 90%. The frequency cosine, Overlap, and TF-IDF cosine distances exhibit the second-best results (except at recall level 100%). Boolean Hamming steadily loses ground, while frequency Hamming is able to keep up with Euclidean. The frequency Hamming crosses Euclidean at recall level 60% in both the PR and FR curves.

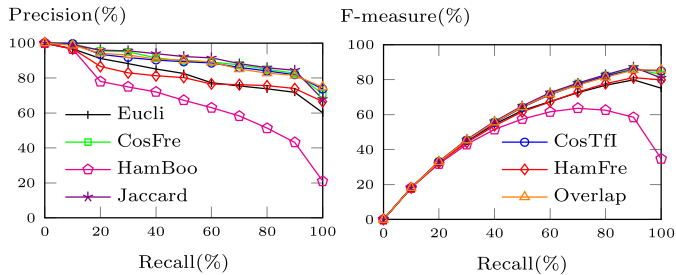

Fig. 14: Average PR and FR curves (4-neighbourhoods)

Overall, in terms of retrieval performance, the results we get are roughly comparable to those we obtained previously (Helmer and Ngo 2015). The main differences are a larger data set, giving the new results more weight, and an improvement for the TF-IDF cosine measure, which is due to fixing a bug in its implementation. As we will see in the following section, though, in terms of cluster performance, we were able to improve considerably.

## 7.3 Clustering Performance

In order to keep the diagrams readable, we restrict ourselves to the following similarity measures in this section: Jaccard, Overlap, and cosine (both TF-IDF and frequency). Similar to their performance in the retrieval case, the cluster performance of the other measures, Hamming (both Boolean and frequency) and Euclidean, is clearly inferior. We make one exception for Ward's criterion, which relies on the Euclidean distance.

However, before comparing the two clustering techniques, HAC and K-means, we need to calibrate their parameters. As already done for the retrieval performance, we have to determine the size of the neighbourhoods for which the clustering algorithms perform well. Additionally, for K-means we need to set a value for $max$, the maximum number of iterations: it turns out that $max = 5$ is a good value. In contrast to the retrieval performance, the best value for $k$ for the clustering algorithms is not as clear-cut, but depends on the employed similarity measure. For TF-IDF cosine, $k = 2$ performs well, for frequency cosine $k$ should be set to 3, while for Jaccard and Overlap, $k = 4$ is the best value. This holds for both clustering approaches, HAC and K-means. For Ward's criterion with the Euclidean distance as used in HAC, $k = 3$ is a good value. For more details on the parameter setup, see Appendix 10.2.

In Figures 15, 16 and 17, we look at purity, NMI, Rand index, precision, recall, and F-measure values for the clustering algorithms using different similarity measures and cluster distance criteria (for HAC). We use the values for $k$ as mentioned above: $k = 3$ for Ward's criterion and for the other cases of HAC and K-means, we set $k$ to 2 for the TF-IDF cosine, to $k$ to 3 for the frequency cosine and to 4 for all other distance measures.

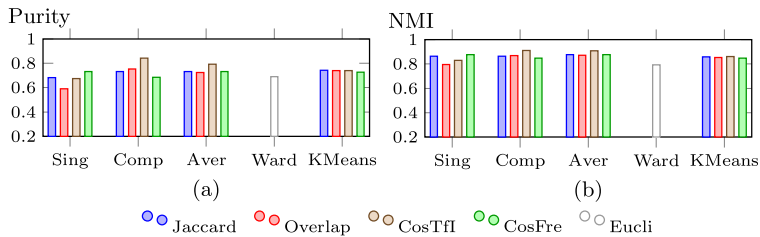

Fig. 15: Purity and NMI for HAC and K-means

For purity and NMI (Figures 15), we see a very similar behaviour, the main difference being the larger values for NMI. We make a couple of interesting observations here. Comparing single-linkage (Sing), complete-linkage (Comp), and average-linkage (Aver), we see that overall single-linkage is inferior to the others, mainly due to the weak performance of the overlap coefficient. This does not come as a surprise, as single-linkage, which only looks at the minimal distance between objects in clusters, has a tendency to create long drawn-out chains. Complete-linkage, considering the maximum distance between objects, avoids this, producing compact clusters of approximately equal diameters. It can be susceptible to outliers, that is why average-linkage is usually preferred. However, in our scenario, this does not seem to be the case, as it outperforms average-linkage. On average, K-means can keep up with single-linkage and average-linkage, but there is a clear winner in the form of HAC using complete-linkage with the TF-IDF cosine measure. It has a purity of 0.842 and an NMI of 0.912. Ward's criterion is only able to outperform single-linkage with Overlap.

Although the values for the Rand index are all very high for the different combinations (see Figure 16(a)), the relative positions do not change compared to the numbers for purity and NMI. The winner is HAC with complete-linkage and the TF-IDF cosine measure again, reaching a value of 0.976. However, given the high number of true negatives, it is not too difficult to achieve a good performance for the Rand index. Thus, we look at the more meaningful measures precision, recall, and F-measure next.

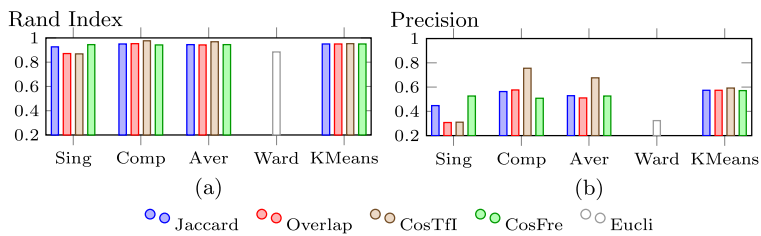

Fig. 16: Rand index and Precision for HAC and K-means

For precision, which is displayed in Figure 16(b), there are no significant changes in terms of relative positioning. Nevertheless, the differences between the various methods become much more distinct. HAC using single-linkage with Overlap and TF-IDF cosine drops to a rather low level of around 0.3, whereas Ward's criterion performs slightly better. HAC with complete-linkage and TF-IDF cosine still shows the strongest performance with a precision of 0.755. The other variants can be found somewhere in between.

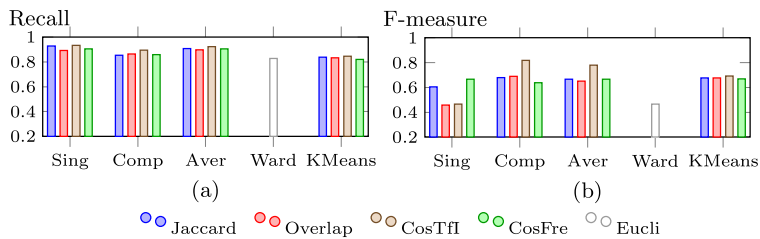

Fig. 17: Recall and F-measure for HAC and K-means

For the first time, we see a significant change in the relative positioning of the different methods for recall (see Figure 17(a)). Overall, HAC with complete-linkage is slightly outperformed by HAC with single-linkage and average-linkage, although it still holds its ground against K-means. However, the higher numbers for the recall come at a price: lower numbers for precision, meaning that clusters with a larger number of false positives are created.

This is actually the motivation for the F-measure, which considers the performance for precision and recall in a balanced way. The results for the F-measure are shown in Figure 17(b), in which a familiar picture re-emerges. The variant of HAC combining complete-linkage with TF-IDF cosine is back on top undoubtedly with an F-measure of 0.819. The relative positioning of the other variants also looks very similar to the one for precision.

In summary, HAC with complete-linkage and the TF-IDF cosine measure shows the strongest performance. Even though it is not the top performer for recall, the differences are rather small and it compensates this with a much better showing for precision. Compared to the previous results (Helmer and Ngo 2015), overall we were able to improve the performance. The Rand index went up from 0.938 to 0.976 (purity and NMI were not used in (Helmer and Ngo 2015)). Although the recall dipped slightly from 0.922 to 0.894, this is still a high value and was more than compensated for by a jump in precision from 0.577 to 0.755 and a subsequent increase in the F-measure from 0.71 to 0.819. The clusters we find with the improved techniques are much more accurate and balanced. Nevertheless, due to the limitations of the current dataset – it is relatively small and balanced – we think further investigations are needed to confirm the results.

## 8 Conclusion and Future Work

We developed a technique based on hypergraphs to represent textiles using a crossing of two threads as the basic building block. Decomposing such a graph into substructures called $k$-neighbourhoods allows us to determine the similarity of the patterns created by the interwoven threads. In turn, this makes it possible to search a collection of textile patterns given a query pattern. We implemented our approach using different distance measures for computing the similarity between multisets of $k$-neighbourhoods. In an experimental evaluation using a data set consisting of 1,600 textile samples, we show that our structural similarity measure can be implemented efficiently and shows very good retrieval and excellent clustering performance. For retrieval, the combination of $k$-neighbourhoods with the TF-IDF cosine and Jaccard distance measure showed very good results, while for clustering, hierarchical agglomerative clustering (HAC) and the TD-IDF cosine measure gave the best results. We note that the experimental results are not so much about improving on an existing approach, but validating the results of our previous work with an extended evaluation, utilising a larger and more diverse data set. We are able to show that the earlier conclusions and insights still hold up, even under different scenarios with new distance and quality measures. As already indicated in the previous section, further evaluation with (very) large and diverse datasets is still needed to gather conclusive evidence. This motivated the automatic or at least semi-automatic generation of datasets as an important task for future work (see below).

For future work, we would like to pursue several goals. First, we would like to investigate further distance measures and variations of *k*-neighbourhoods to identify ways to improve our textile similarity measure. At the moment the modelling of the textiles used for the hypergraph representation has to largely be done manually. In order to automate this process, image-processing techniques for extracting a thread structure and mapping it to graphs would be an interesting topic to look into. This would facilitate the construction of a gold standard data set that can be used to stress-test the behaviour, accuracy, and robustness of the proposed approach using many different textile patterns. The application of deep learning algorithms may also be a promising direction to take, followed by a comparison of such a technique to our similarity measure. We also apply data warehouse to store textiles to reduce the execution time (Ngo et al. 2019, 2020) or knowledge base about textile to improve the performances (Ngo and Cao 2011) .

## 9 Acknowledgements

This research is an extension of Helmer and Ngo (2015), which was funded by the SimTex project (IN5029) at the Free University of Bozen-Bolzano. It is part of the CONSUS research program which is funded under the SFI Strategic Partnerships Programme (16/SPP/3296) and is co-funded by Origin Enterprises Plc. Dr. Vuong M. Ngo implemented the primary part of this research as he worked in University College Dublin, Ireland.

## 10 Appendices

10.1 Vietnamese Textile Patterns

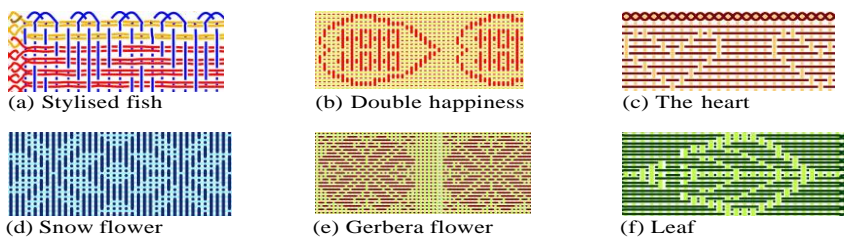

(a) Stylised fish    (b) Double happiness    (c) The heart

(d) Snow flower    (e) Gerbera flower    (f) Leaf

Fig. 18: More examples of Vietnam weaving patterns

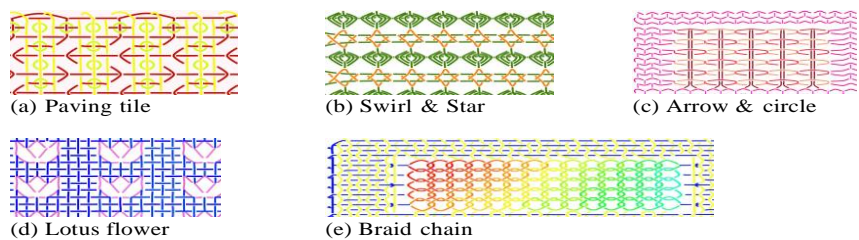

(a) Paving tile    (b) Swirl & Star    (c) Arrow & circle

(d) Lotus flower    (e) Braid chain

Fig. 19: More examples of Vietnam mix-fabric patterns

10.2 Parameter Setup

*10.2.1 Maximum Number of Iterations for K-Means*

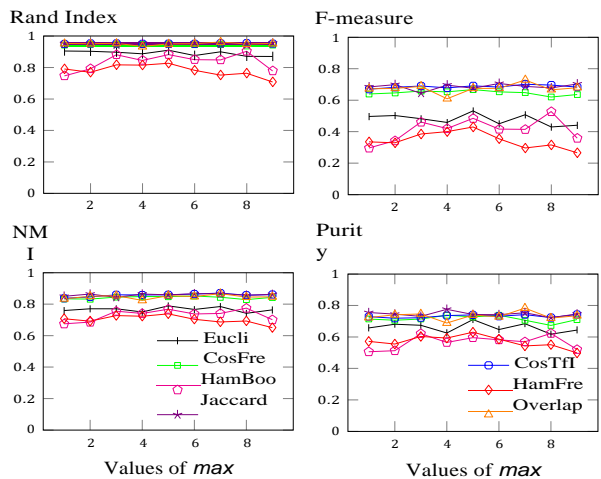

Fig. 20: Determining the maximum number of iterations for K-means

Figure 20 shows the performance for K-means (for different quality measures) when varying the maximum number of iterations. Here we have included the Euclidean and Hamming distances as well to verify their weaker performance. The size of the neighbourhood, $k$, was set to 2 for TF-IDF cosine, to 3 for frequency cosine, and to 4 for all other distance measures. (We will look at the impact of $k$ in just a moment in the following section.) As can be seen in Figure 20, K-means stabilises quite quickly. There are no significant improvements beyond $max = 5$. On the contrary, some distance measures are even slightly worse for higher values.

*10.2.2 Size of Neighbourhood*

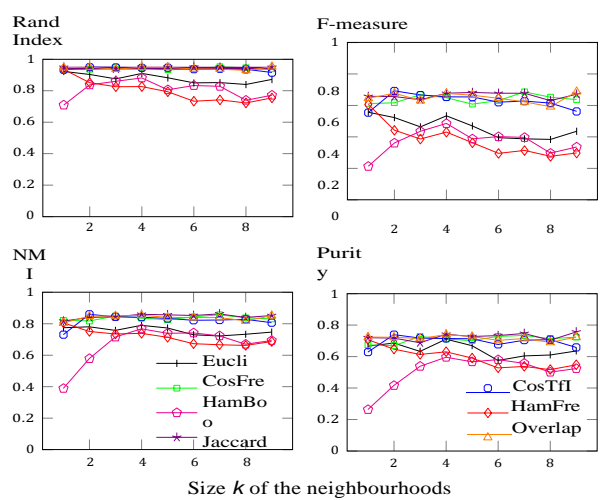

Fig. 21: Determining the size of the neighbourhood for K-means

Figure 21 shows the performance for K-means for different similarity measures and neighbourhood sizes (the maximum number of iterations, $max$, is set to 5). The Euclidean and Hamming distances are included for the sake of completeness, their performance is clearly inferior to the other distance measures. The TF-IDF cosine measure has a clear peak for $k = 2$ for all the different quality measures. For frequency cosine, the peak is shifted by one position, moving to $k = 3$. In the case of Overlap, the peak is shifted even further to $k = 4$. For Jaccard, $k = 4$ is the best value on average, only its recall performance is fractionally better for $k = 5$.

For the figures describing the performance of HAC, we drop the Euclidean and Hamming distances (except for Ward's criterion, which uses Euclidean) to make them more readable. Figure 22 shows the results for the cosine measures when varying the size $k$ of the neighbourhood. The superiority of complete-linkage with TF-IDF cosine can be clearly seen, even for values different from $k = 2$. Nevertheless, it performs best for $k = 2$; this is also the case for average-linkage with TF-IDF cosine. Single-linkage looks slightly different, but as it

38                                                                                    Vuong M. Ngo et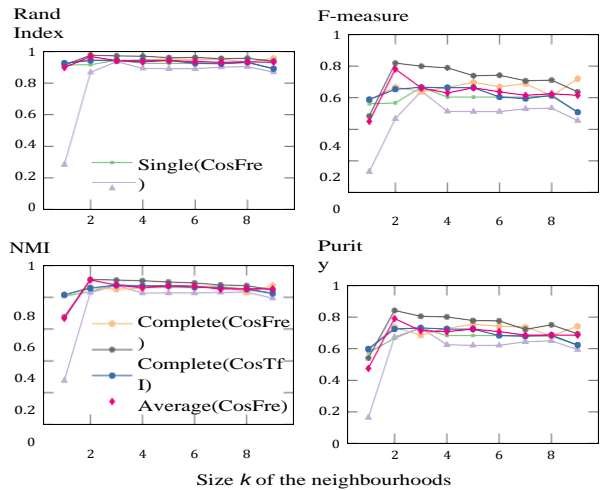

Fig. 22: Determining $k$ for HAC using the cosine measures

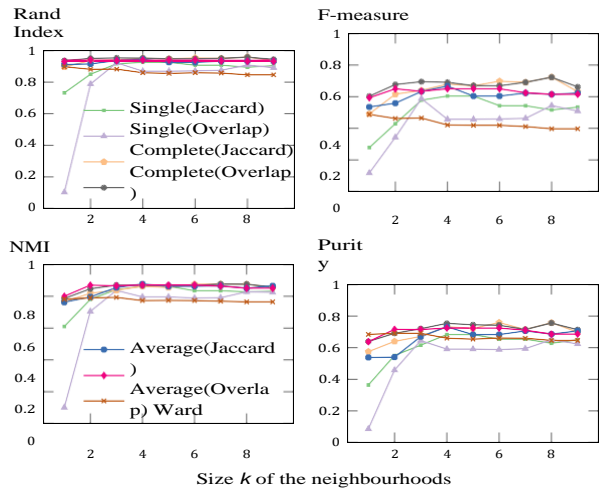

Fig. 23: Determining $k$ for HAC using other distance measures

is outperformed by all the other combinations, it is not relevant anyway. On average, frequency cosine performs best for $k = 3$.

Finally, we come to the remaining distance measures for HAC, Jaccard and Overlap, whose performance, along with that for Ward's criterion, is shown in Figure 23. Complete-linkage with Jaccard and Overlap exhibits the strongest performance, peaking at $k = 4$. Average-linkage shows a similar picture, although at a lower level of performance. As in the previous figure, the behaviour of single-linkage is not that clear, but as it brings up the rear with Ward's criterion, it should not be chosen anyway.